\documentclass{TTP_DSL2006}
\usepackage[english,russian]{babel} 
\usepackage{array}
\usepackage{mathtext}
\usepackage{textcomp}
\usepackage{cite}
\usepackage{graphicx,wrapfig}
\usepackage{sidecap,caption}    
\sidecaptionvpos{figure}{c}     

\newcommand{\be}{\begin{equation}}
\newcommand{\ee}{\end{equation}}
\newcommand{\ve}{\varepsilon}

\usepackage{graphicx}
\usepackage[intlimits]{amsmath}
\usepackage{amssymb}
\usepackage{subfig}     
\usepackage{exscale}    
\usepackage{hyperref}

\newcommand{\titleformat}{\sffamily\bfseries \large}				
\newcommand{\authorformat}{\sffamily \large}						
\newcommand{\keywordsformat}{\noindent \small \sffamily}			
\newcommand{\abstractformat}{\noindent \textbf}						
\newcommand{\contentformat}{\rmfamily \normalsize\vspace{18pt}}		
\newcommand{\email}{\sffamily \small \vspace{-8pt}}					
\renewcommand{\subsection}{\textbf}									

\hypersetup{
    colorlinks,%
    citecolor=black,%
    filecolor=black,%
    linkcolor=black,%
    urlcolor=black
}

\begin{document}

\selectlanguage{english} 

\title{\titleformat Effect of mechanical stresses on coercive force and 
                    saturation remanence of ensemble of 
                    dual-phase interacting nanoparticles}

\author{\authorformat Leonid Afremov \inst{1} and Yury Kirienko \inst{2}$^{,*}$}

\institute{\sffamily School of Natural Science, Far-Eastern Federal University, Vladivostok, Russia}
\maketitle

\begin{center}
\email{$^{1\,}$afremovl@mail.dvgu.ru, %
       $^{2\,}$yury.kirienko@gmail.com, %
       $^{*}$corresponding author}
\end{center}

\vspace{2mm} \hspace{-7.7mm} \normalsize 
\keywordsformat{\textbf{Keywords:} heterophase particles, mechanical stress, magnetic states, 
 elongated nanoparticles, coatings, interfacial exchange interaction, saturation remanence,
 maghemite, cobalt.}

\contentformat

\abstractformat{Abstract.} 
In the dual-phase model of interacting nanoparticles stretching leads to a decrease in both 
coercive force $H_c$ 
and saturation remanence $I_{rs}$, and compression — to their growth. 
Magnetostatic interaction between particles also decreases both $H_c$ and $I_{rs}$.
Theoretical analysis was carried out in the framework of the dual-phase system of interacting particles 
on the example of nanoparticles $\gamma$-$Fe_2O_3$, epitaxially coated with cobalt.

\section{Introduction}
An important factor affecting the process of magnetization of an ensemble of small (single-domain) 
particles is their magnetostatic interaction.
For example, consider a system of $N$ magnetic grains of size $a$, 
uniformly distributed in a nonmagnetic matrix. The field produced by each of these particles on the neighboring 
can be estimated as $H_m \sim 2 I_s N a^3/R^3 = 2 c I_s$, where $R$ — the average distance between particles,
$c=Na^3/R^3$ — volume concentration of magnetic material in the system.
In the materials for magnetic recording, magnetics with high values of spontaneous magnetization
($I_s\sim350-500$~G) is commonly used.
Therefore, at concentrations $c\sim0.1-0.2$, magnetostatic interaction may have significant influence 
on magnetization of the low-coercive particles.

To analyze the effect of magnetic interaction on the magnetization of an ensemble of 
chemically inhomogeneous nanoparticles, we use the model of dual-phase particles, 
described in detail in \cite{Afremov2012}:

\begin{figure}
    \centering
    \includegraphics{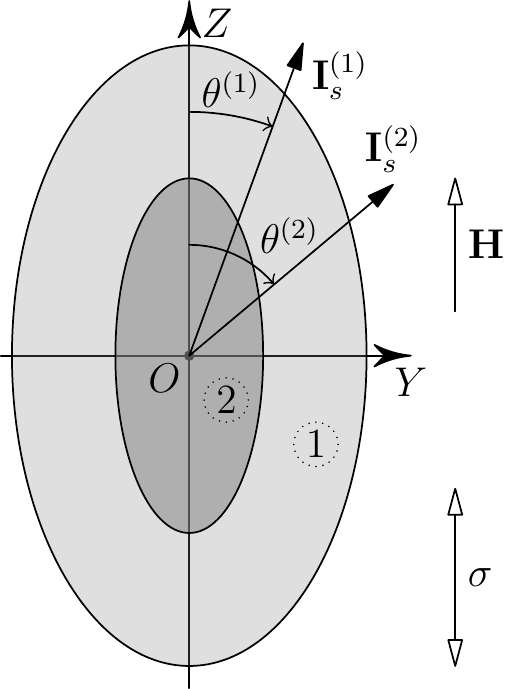}
    \caption{\small Illustration to the model\\ of dual-phase particle}
    \label{fig:1}
\end{figure}
\section{Model}
\begin{enumerate}
\item  Homogeneously magnetized nanoparticle (phase {\em 1}) of volume $V$
has the form of ellipsoid with elongation\footnote{{\em Elongation} — 
the ratio of the length $a$ of semi-major axis of the ellipsoid to the length $b$ of semi-minor one.}  $q_1$, 
and its long axis is oriented along the $Oz$-axis (see Fig.~\ref{fig:1}).

\item Nanoparticle contains an uniformly magnetized ellipsoidal inclusion (phase {\em 2}) 
with a volume $v=\varepsilon V$ and elongation $q$.

\item It is considered that the axes of crystallographic anisotropy of both uniaxial ferromagnets are parallel 
to the long axes of the ellipsoids, and the vectors of spontaneous magnetization of 
phases ${{\mathbf I}}^{(1)}_s$ and ${{\mathbf I}}^{(2)}_s$ lie in the plane $yOz$, 
that contains the long axes of the magnetic phases, and make angles $\theta^{(1)}$ and $\theta^{(2)}$ 
with the $Oz$ axis, respectively.

\item Both external magnetic field $H$ and uniaxial mechanical stresses $\sigma $ are applied along the $Oz$-axis. 

\item The volume of nanoparticles exceeds the volume of superparamagnetic transition.
In other words, thermal fluctuations are not taken into account.

\end{enumerate}

The calculation of fields of magnetostatic interaction between dual-phase particles was performed 
using the method of the random field interaction \cite{Afremov1996,Belokon2001}.
It was found that at low concentrations of magnetic nanoparticles in a nonmagnetic matrix 
($c<0.1$) the distribution function of the projection of the field interaction 
on the direction of the external magnetic field $\vec{H}$ is described by the Cauchy law:
\be\label{eq:1}
W\left(h\right)=\frac{1}{\pi}\frac{B}{B^2+{(h-\alpha I)}^2}, 
\ee
where $\alpha = 8\pi/(5-\widehat{N})$, $\widehat{N}$ — the demagnetization factor. 
Parameter $B$ and magnetization $I$ are determined by the system of equations:
\be\label{eq:2}
\left\{
  \begin{aligned}
    B&=&5c\int\left\{(N_1+N_3)\left[I_{s1}(1-\ve)+I_{s2}\,\ve\right]+
    (N_2+N_4)\left|I_{s1}(1-\ve)-I_{s2}\,\ve\right|%
    \right\}W\left(h\right)dh,\\
%
    I&=&c\int\left\{(N_1-N_3)\left[I_{s1}(1-\ve)+I_{s2}\,\ve\right]+
    (N_2-N_4)\left[I_{s1}(1-\ve)-I_{s2}\,\ve\right]%
    \right\}W\left(h\right)dh,
  \end{aligned}%
\right.
\ee
where $\ve=v/V$ — relative volume of the second phase (inclusion),
$V$ — volume of nanoparticle, $v$ — volume of inclusion.
According to~\cite{Afremov2012, Afremov2010}, $N_1$ is the number of particles, 
whose magnetic moments of both phases are parallel  to the external field $\vec{H}$; 
$N_2$  is the number of particles with the magnetic moments of the first phase that are parallel to $\vec{H}$
and the magnetic moments of the second phase are antiparallel to $\vec{H}$.
(If we consider  $\vec{H}$ to be directed «upward», then we can denote the first state as «$\uparrow\uparrow$»,
the second one as «$\uparrow\downarrow$», the third one as «$\downarrow\downarrow$».
and the last one (4th) as «$\downarrow\uparrow$». 
The corresponding numbers of particles in each of these states we denote as $N_1$, $N_2$, $N_3$ and $N_4$.)

In the case of large concentrations ($c>0.1$), the distribution of the random field interaction obeys the normal law:
\be\label{eq:3}
W\left(h\right)
    =\frac{1}{\sqrt{2\pi}B}\exp\left\{-\frac{{\left(h-\widetilde{\alpha} I\right)}^2}{2B^2}\right\}, 
\ee
where
\be\label{eq:4}
    B=c\sqrt{\int\left\{(N_1{+}N_3)\left[I_{s1}(1-\ve)+I_{s2}\ve\right]+
    (N_2{+}N_4)\left|I_{s1}(1-\ve)-I_{s2}\ve\right|%
    \right\}^2W\left(h\right)dh},
\ee
and the magnetization $I$ is given by the same equation as in~\eqref{eq:2}.

The obtained self-consistent equations~\eqref{eq:1}\,—\,\eqref{eq:4} allow us to estimate 
the characteristic field interaction $B$, and to calculate the magnetization $I$ 
of the system of dual-phase interacting nanoparticles.

\begin{wrapfigure}{l}{0.47\textwidth}
    \includegraphics[scale=0.9]{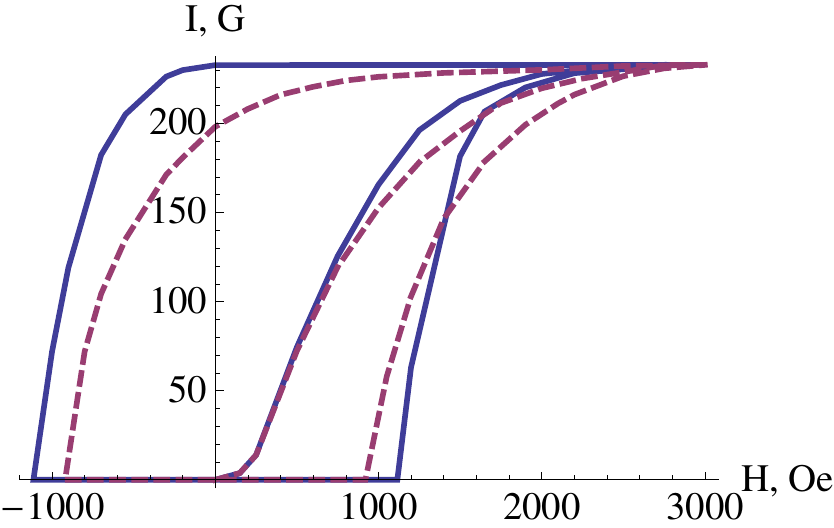}
    \caption{\small Hysteresis loops and magnetization curves of ensemble of non-interacting
     ({\em solid curve}, $c=0$) 
    and interacting ({\em dashed curve}, $c=0.24$) spherical nanoparticles with the following characteristics: 
    $q_1=1,\  A_{in}=0,\ k_{\sigma }=0$}
\label{fig:3.2.1}
\end{wrapfigure}

\begin{wrapfigure}{r}{0.47\textwidth}
    \vspace{-9.5cm}
    \includegraphics[scale=0.9]{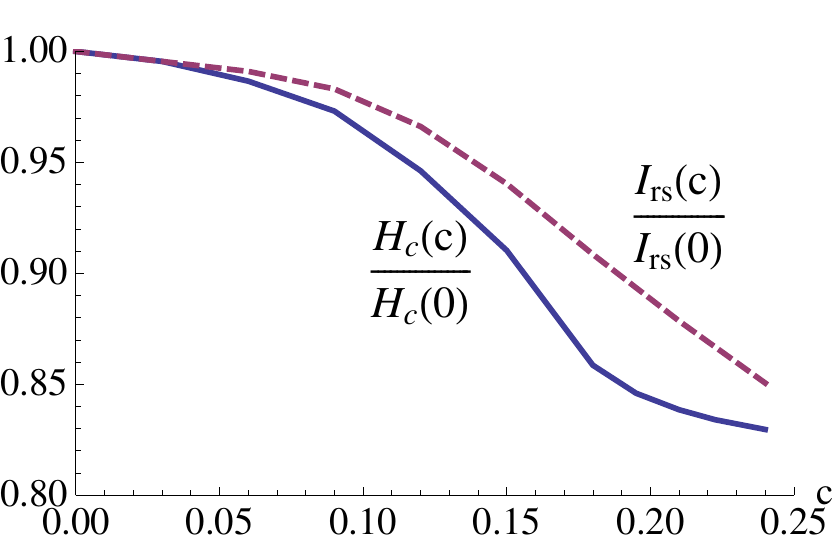}
    \caption{\small Relative coercive force $H_c(c)/H_c(0)$ and 
             relative saturation remanence $I_{rs}(c)/I_{rs}(0)$ 
             on the volume concentration $c$ of magnetic nanoparticles}
    \label{fig:3.2.2}
\end{wrapfigure}

\section{Results and discussion}
The influence of magnetic interaction on the hysteresis loop and magnetization curve 
of the ensemble of nanoparticles $CoFe_2O_4$ — $\gamma$-$Fe_2O_3$ is shown in Fig.~\ref{fig:3.2.1}.
When the volume concentration of magnetic particles becomes significant\footnote{%
    Here and below, a record «$c = 0$» means that the concentration is small, 
    but nonzero (otherwise the magnetization of the sample would also turn to zero).}
(e.\,g., $c=24\,\%$, see Fig.~\ref{fig:3.2.1}), 
the magnetostatic interaction smooths the hysteresis loop and lowers the magnetization curve.
The interaction between the particles decreases both the coercive force $H_c$ (from 1115\,Oe to 920\,Oe) 
and saturation remanence $I_{rs}$ (from 232\,G to 198\,G).
Moreover, the hysteresis characteristics vary non-linearly as a function of the magnetic interaction, 
and for small values of the volume concentration of magnetic particles (say, equal to $c = 9\,\%$) 
are reduced by less than 3\,\% (see Fig.~\ref{fig:3.2.2}).

The reduction of hysteresis characteristics with growth of concentration of magnetic particles 
is related to disarranging influence of the magnetic interaction on the distribution of magnetic moments.

\begin{figure}
    \centering
    \subfloat[]{\includegraphics{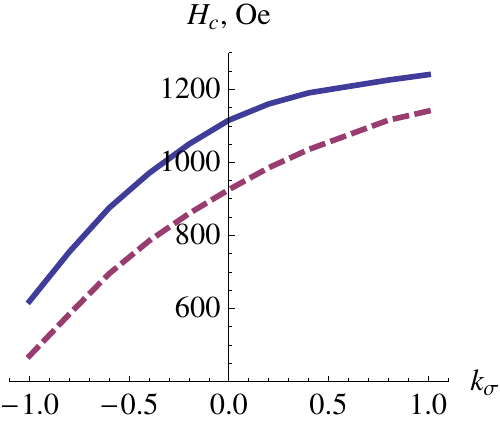}}
    \subfloat[]{\includegraphics{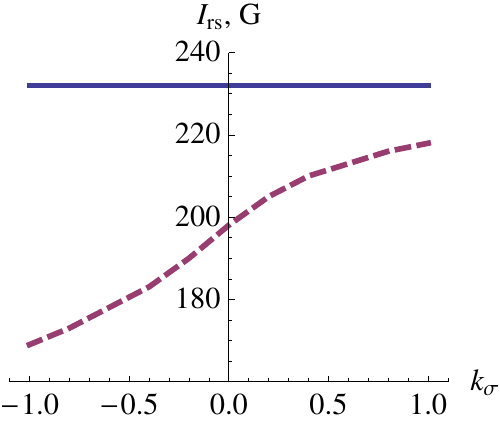}}
    \subfloat[]{\includegraphics{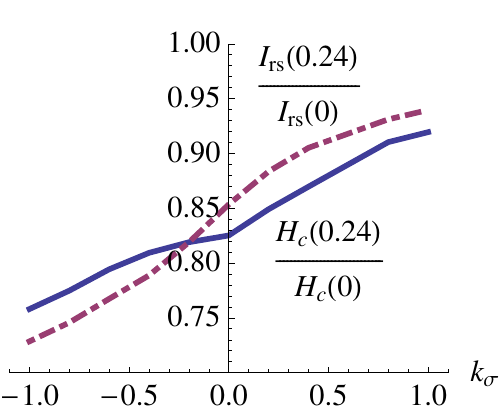}}
    \caption
    {\small (a) the dependence of the coercive force  $H_c$ and (b) saturation remanence  $I_{rs}$  
    of an ensemble of non-interacting ({\em solid curves}) and interacting ({\em dashed curves}) spherical nanoparticles  
    on the relative stress  $k_{\sigma }$;
    (c) the dependence of the relative coercive force $H_c (c = 0.24)/H_c (c = 0)$ 
    and the relative saturation remanence $I_{rs} (c = 0.24) / I_{rs} (c = 0)$ 
    on the mechanical stresses; $q_1=1,\ A_{in}=0$
    }
    \label{fig:4}
\end{figure}

Fig.~\ref{fig:4} shows the dependence of the coercive force $H_c$ and 
saturation remanence $I_rs$ on the mechanical stresses $\sigma$,
which are determined by the dimensionless constant $k_{\sigma}=\lambda\sigma/K_A$
(where $\lambda$ — magnetostriction constant, $K_A$ — constant of crystallographic anisotropy).
As expected, the stretching decreases the coercive force and the compression increases it, 
both in an ensemble of non-interacting nanoparticles and interacting nanoparticles (see Fig.~\ref{fig:4}a).
This can be explained by the fall of the critical fields of magnetization reversal 
under tension and their growth under compression~\cite{Afremov2010}. 
The saturation remanence of interacting nanoparticles changes in a similar way, due to the mechanical stresses.
At the same time, the saturation remanence $I_{rs}$ of non-interacting nanoparticles 
is independent on the stresses (see Fig.~\ref{fig:4}b). 
Also note that the hysteresis characteristics of an ensemble of interacting nanoparticles 
under tension vary stronger than in compression.
This behavior of $H_c$ and $I_{rs}$ is connected to the fact that the magnetostatic interaction 
stronger shuffles the magnetic moments of low-coercive nanoparticles, 
and has a smaller influence on the particles in the high-coercive states (see Fig.~\ref{fig:4}c).

\section{Acknowledgments}
The work was partly supported by grants of Ministry of Education and Science: 
Federal Contract \No\,02.740.11.0549,  
Federal Contract \No 14.740.11.0289 
and
Federal Contract \No 07.514.11.4013.

\bibliographystyle{ieeetr}
\bibliography{2_guangzhou}
\end{document}